\documentclass[longauth]{aa} 

\usepackage{graphicx}
\usepackage{txfonts}
\usepackage{multirow}
\usepackage{colortbl}

\usepackage{hyperref}  
\hypersetup{colorlinks=true,linkcolor=[rgb]{1.,0.2,0.2},citecolor=[rgb]{0.1,0.4,1.},filecolor=[rgb]{0.7,0.2,0.2},urlcolor=[rgb]{0.7,0.2,0.2}}

\usepackage{color}
\definecolor{blue}{rgb}{0., 0., 1}

\newcommand{\hii}{\textrm{H}\textsc{ii}}

\newcommand{\oiii}{[\textrm{O}\textsc{iii}]}

\newcommand{\oiidoublam}{[\textrm{O}\textsc{ii}]\ensuremath{\lambda3727,3729}}

\newcommand{\oiiiv}{[\textrm{O}\textsc{iii}]\ensuremath{\lambda5007}}

\newcommand{\oiiidoublam}{[\textrm{O}\textsc{iii}]\ensuremath{\lambda\lambda4959,5007}}
 
\newcommand{\ha}{\ifmmode {\rm H}\alpha \else H$\alpha$\fi}
\newcommand{\halam}{\ifmmode {\rm H}\alpha \lambda6563 \else H$\alpha$ $\lambda$6563 \fi}
\newcommand{\hb}{\ifmmode {\rm H}\beta \else H$\beta$\fi}
\newcommand{\hg}{\ifmmode {\rm H}\gamma \else H$\gamma$\fi}
\newcommand{\hblam}{\ifmmode {\rm H}\beta \lambda4861 \else H$\beta$ $\lambda$4861 \fi}
\newcommand{\lya}{\ifmmode {\rm Ly}\alpha \else Ly$\alpha$\fi}
\newcommand{\pg}{\ifmmode {\rm P}\gamma \else Pa$\gamma$\fi}
\newcommand{\lyb}{\ifmmode {\rm Ly}\beta \else Ly$\beta$\fi}
\newcommand{\lyg}{\ifmmode {\rm Ly}\gamma \else Ly$\gamma$\fi}

\newcommand{\neiii}{\textrm{Ne}\textsc{iii}]\ensuremath{\lambda3869}}

\newcommand{\flyc}{\ifmmode  \mathrm{f}_\mathrm{esc}\mathrm{(LyC)} \else $\mathrm{f}_\mathrm{esc}\mathrm{(LyC)}$\fi}

\def\ergs{\ifmmode \mathrm{erg\hspace{1mm}s}^{-1} \else erg s$^{-1}$\fi}

\def\micron{\ifmmode \mu\mathrm{m} \else $\mu$m\fi}
\def\msun{\ifmmode \mathrm{M}_{\odot} \else M$_{\odot}$\fi}
\def\msunyr{\ifmmode \mathrm{M}_{\odot} \hspace{1mm}{\rm yr}^{-1} \else $\mathrm{M}_{\odot}$ yr$^{-1}$\fi}
\def\zsun{\ifmmode Z_{\odot} \else Z$_{\odot}$\fi}
\def\lsun{\ifmmode L_{\odot} \else L$_{\odot}$\fi}
\def\mstar{\ifmmode \mathrm{M}_{\star} \else M$_{\star}$\fi}
\newcommand{\JWST}{\textit{JWST}}
\newcommand{\HST}{\textit{HST}}

\usepackage{orcidlink}

\newcommand{\orcid}[1]{\href{https://orcid.org/#1}{\textcolor[HTML]{A6CE39}{\aiOrcid}}}

\begin{document}

\titlerunning{\JWST\ probes super-faint low-metallicity stellar ionizers at $z\simeq6.14$}

\title{Extreme Ionising Properties of Metal-Poor, M$_{\rm UV} \simeq -12$ Star Complex in the first Gyr\thanks{Based on observations collected with the James Webb Space Telescope (\JWST) and Hubble Space Telescope (\HST). 
These observations are associated with \JWST\ GO program n.1908 (PI E. Vanzella), GTO n.1208 (CANUCS, PI C. Willott) and GTO n.1176 (PEARLS, PI R. Windhorst).}}

\authorrunning{Eros Vanzella et al.}
\author{
E.~Vanzella\inst{\ref{inafbo}}\fnmsep\thanks{E-mail: \href{mailto:eros.vanzella@inaf.it}{eros.vanzella@inaf.it}}$^{\orcidlink{0000-0002-5057-135X}}$ \and
F.~Loiacono\inst{\ref{inafbo}}$^{\orcidlink{0000-0002-8858-6784}}$ \and
M.~Messa\inst{\ref{inafbo}}$^{\orcidlink{0000-0003-1427-2456}}$ \and
M.~Castellano\inst{\ref{inafroma}}$^{\orcidlink{0000-0001-9875-8263}}$ \and
P.~Bergamini \inst{\ref{unimi},\ref{inafbo}}$^{\orcidlink{0000-0003-1383-9414}}$ \and
A.~Zanella \inst{\ref{inafpd}}$^{\orcidlink{0000-0001-8600-7008}}$\and
F.~Annibali\inst{\ref{inafbo}}$^{\orcidlink{0000-0003-3758-4516}}$\and
B.~Sun\inst{\ref{UniMiss}}$^{\orcidlink{0000-0001-7957-6202}}$\and
M.~Dickinson\inst{\ref{tucson}}$^{\orcidlink{0000-0001-5414-5131}}$\and
A.~Adamo\inst{\ref{univstock}}$^{\orcidlink{0000-0002-8192-8091}}$\and
F.~Calura\inst{\ref{inafbo}}$^{\orcidlink{0000-0002-6175-0871}}$\and
M.~Ricotti\inst{\ref{univmaryland}}$^{\orcidlink{0000-0003-4223-7324}}$\and
P.~Rosati \inst{\ref{unife},\ref{inafbo}}$^{\orcidlink{0000-0002-6813-0632}}$\and
M.~Meneghetti \inst{\ref{inafbo}}$^{\orcidlink{0000-0003-1225-7084}}$\and
C.~Grillo \inst{\ref{unimi},\ref{inafiasf}}$^{\orcidlink{0000-0002-5926-7143}}$\and
M.~Brada{\v c}\inst{\ref{uniLjubljana},\ref{unicalifornia}}$^{\orcidlink{0000-0001-5984-0395}}$\and 
C.~J. Conselice\inst{\ref{Manchester}}$^{\orcidlink{0000-0003-1949-7638}}$\and
H.~Yan\inst{\ref{UniMiss}}$^{\orcidlink{0000-0001-7592-7714}}$\and 
A.~Bolamperti\inst{\ref{unipd},\ref{inafpd}}$^{\orcidlink{0000-0001-5976-9728}}$\and
U.~Me\v{s}tri\'{c} \inst{\ref{unimi},\ref{inafbo}}$^{\orcidlink{0000-0002-0441-8629}}$\and
R.~Gilli\inst{\ref{inafbo}}$^{\orcidlink{0000-0001-8121-6177}}$\and
M.~Gronke\inst{\ref{maxplanck}}$^{\orcidlink{0000-0003-2491-060X}}$\and
C.~Willott\inst{\ref{nrccanada}}$^{\orcidlink{0000-0002-4201-7367}}$\and
E.~Sani\inst{\ref{esochile}}$^{\orcidlink{0000-0002-3140-4070}}$\and
A.~Acebron\inst{\ref{unimi}}$^{\orcidlink{0000-0003-3108-9039}}$\and
A.~Comastri\inst{\ref{inafbo}}$^{\orcidlink{0000-0003-3451-9970}}$\and
M.~Mignoli\inst{\ref{inafbo}}$^{\orcidlink{0000-0002-9087-2835}}$\and
C.~Gruppioni\inst{\ref{inafbo}}$^{\orcidlink{0000-0002-5836-4056}}$\and
A.~Mercurio\inst{\ref{univsalerno},\ref{inafnapoli}}$^{\orcidlink{0000-0001-9261-7849}}$\and
V.~Strait\inst{\ref{dawn},\ref{univcope}}$^{\orcidlink{0000-0002-6338-7295}}$\and
R.~Pascale\inst{\ref{inafbo}}$^{\orcidlink{0000-0002-6389-6268}}$\and
M.~Annunziatella\inst{\ref{spain_astrob}}$^{\orcidlink{0000-0002-8053-8040}}$\and
B.~L.~Frye\inst{\ref{UAZ}}$^{\orcidlink{0000-0003-1625-8009}}$\and
L.~D.~Bradley\inst{\ref{stsci}}$^{\orcidlink{0000-0002-7908-9284}}$\and
N.~A.~Grogin\inst{\ref{stsci}}$^{\orcidlink{0000-0001-9440-8872}}$\and
A.~M.~Koekemoer\inst{\ref{stsci}}$^{\orcidlink{0000-0002-6610-2048}}$\and
S.~Ravindranath\inst{\ref{Goddard}\and \ref{CatholicUnivAmerica}}$^{\orcidlink{0000-0002-5269-6527}}$\and
J.~C.~J.~D’Silva\inst{\ref{ICRAR}}$^{\orcidlink{0000-0002-9816-1931}}$\and
J.~Summers\inst{\ref{ArizonaSU}}$^{\orcidlink{0000-0002-7265-7920}}$\and
G.~Rihtar{\v s}i{\v c}\inst{\ref{uniLjubljana}}$^{\orcidlink{0009-0009-4388-898X}}$ \and
R.~Windhorst\inst{\ref{ASU},\ref{CatholicUnivAmerica}}$^{\orcidlink{0000-0001-8156-6281}}$
}
\institute{
INAF -- OAS, Osservatorio di Astrofisica e Scienza dello Spazio di Bologna, via Gobetti 93/3, I-40129 Bologna, Italy \label{inafbo} 
\and
INAF -- Osservatorio Astronomico di Roma, Via Frascati 33, 00078 Monteporzio Catone, Rome, Italy\label{inafroma}
\and
Dipartimento di Fisica, Università degli Studi di Milano, Via Celoria 16, I-20133 Milano, Italy\label{unimi}
\and
INAF Osservatorio Astronomico di Padova, vicolo dell'Osservatorio 5, 35122 Padova, Italy\label{inafpd}
\and
Department of Physics and Astronomy, University of Missouri, Columbia, MO 65211, USA \label{UniMiss}
\and
NSF's National Optical-Infrared Astronomy Research Laboratory, 950 N. Cherry Ave., Tucson, AZ 85719, USA\label{tucson}
\and
Department of Astronomy, Oskar Klein Centre, Stockholm University, AlbaNova University Centre, SE-106 91, Sweden\label{univstock}
\and
Department of Astronomy, University of Maryland, College Park, 20742, USA\label{univmaryland}
\and
Dipartimento di Fisica e Scienze della Terra, Università degli Studi di Ferrara, Via Saragat 1, I-44122 Ferrara, Italy\label{unife}
\and 
INAF -- IASF Milano, via A. Corti 12, I-20133 Milano, Italy\label{inafiasf}
\and
University of Ljubljana, Department of Mathematics and Physics, Jadranska ulica 19, SI-1000 Ljubljana, Slovenia\label{uniLjubljana}
\and 
Department of Physics and Astronomy, University of California Davis, 1 Shields Avenue, Davis, CA 95616, USA\label{unicalifornia}
\and
  Jodrell Bank Centre for Astrophysics, Alan Turing Building, University of Manchester, Oxford Road, Manchester M13 9PL, UK\label{Manchester}
\and
Dipartimento di Fisica e Astronomia, Università degli Studi di Padova, Vicolo dell'Osservatorio 3, I-35122 Padova, Italy\label{unipd}
\and
Max Planck Institut für Astrophysik, Karl-Schwarzschild-Straße 1, D-85748 Garching bei M\"unchen, Germany\label{maxplanck}
\and
NRC Herzberg, 5071 West Saanich Rd, Victoria, BC V9E 2E7, Canada\label{nrccanada}
\and
European Southern Observatory, Alonso de Córdova 3107, Casilla 19, Santiago 19001, Chile\label{esochile}
\and
Dipartimento di Fisica "E.R. Caianiello," Universit\`a Degli Studi di Salerno, Via Giovanni Paolo II, I-84084 Fisciano (SA), Italy\label{univsalerno}
\and
INAF -- Osservatorio Astronomico di Capodimonte, Via Moiariello 16, I-80131 Napoli, Italy\label{inafnapoli}
\and
Cosmic Dawn Center (DAWN), Denmark\label{dawn}
 \and
Niels Bohr Institute, University of Copenhagen, Jagtvej 128, DK-2200 Copenhagen N, Denmark\label{univcope}
\and
Centro de Astrobiolog\'ia (CAB), CSIC-INTA, Ctra. de Ajalvir km 4, Torrej\`on de Ardoz, E-28850, Madrid, Spain\label{spain_astrob}
\and
Department of Astronomy/Steward Observatory, University of Arizona, 933 N. Cherry Avenue, Tucson, AZ 85721, USA \label{UAZ}
\and 
Space Telescope Science Institute (STScI), 3700 San Martin Drive, Baltimore, MD 21218, USA\label{stsci}
\and
Astrophysics Science Division, NASA Goddard Space Flight Center, 8800 Greenbelt Road, Greenbelt, MD 20771, USA \label{Goddard} 
\and
Center for Research and Exploration in Space Science and Technology II, Department of Physics, Catholic University of America, 620 Michigan Ave N.E., Washington DC 20064, USA \label{CatholicUnivAmerica} 
\and
International Centre for Radio Astronomy Research (ICRAR) and the International Space Centre (ISC), The University of Western Australia, M468, 35 Stirling Highway, Crawley, WA 6009, Australia\label{ICRAR}
\and
School of Earth and Space Exploration, Arizona State University, Tempe, AZ 85287-1404, USA\label{ArizonaSU}
\and
School of Earth and Space Exploration, Arizona State University,
Tempe, AZ 85287-6004, USA \label{ASU}
}

\date{} 

 
\abstract
{
We report the serendipitous discovery of a faint (M$_{\rm UV} > -12.2$), low-metallicity (Z~$\sim 0.02$ Z${\odot}$), ionizing source (dubbed T2c) with a spectroscopic redshift of z$=$6.146.
T2c is part of a larger structure amplified by the Hubble Frontier Field galaxy cluster MACSJ0416, and was observed with \JWST/NIRSpec IFU. Stacking the short-wavelength NIRCam data reveals no stellar continuum detection down to a magnitude limit 
of $m_{\rm UV} \simeq 31.0$ ($3\sigma$). However, prominent \hb, \oiiidoublam, and \ha\ emissions are detected, with equivalent widths exceeding $200$ \AA, $800$ \AA, and $1300$ \AA\ ($3\sigma$), respectively. The corresponding intrinsic (magnification-corrected $\times 23 \pm 3$) ultraviolet and optical rest-frame magnitudes exceed 34.4 and 33.9 (corresponding to M$_{\rm UV}$ and M$_{\rm opt}$ fainter than -12.2 and -12.8, at $\lambda_{rest} \sim 2000$ \AA\ and $\sim 5000$ \AA, respectively), suggesting a stellar mass lower than a few $10^4$~\msun\ under an instantaneous burst scenario. The inferred ionizing photon production efficiency ($\xi_{ion}$) is high, $\xi_{ion} \gtrsim 26.08(25.86)$ $3(5)\sigma$, assuming no dust attenuation and no Lyman continuum leakage, indicating the presence of massive stars despite the low mass of the object. The very poor sampling of the initial mass function at such low mass star-forming complex suggests that the formation of very massive stars might be favored in very low metallicity environments. 
T2c is surrounded by Balmer and weak oxygen emission on a spatial scale of a few hundred parsecs after correcting for lensing effects. This system resembles an \hii\ region potentially powered by currently undetected, extremely efficient, low-metallicity star complexes or clusters.
We propose that massive O-type stars populate this low-mass and metallicity high-redshift satellites, likely caught in an early and short formation phase, contributing to the ionization of the surrounding medium.
}

   \maketitle

\section{Introduction}
\label{sect:intro}





In a scenario in which star-formation is the dominant route for cosmic hydrogen reionization ($z>5$), the role of stellar clusters hosted in high-z galaxies becomes crucial (\citealt{adamo2024}; see also \citealt{katz13, he20_ricotti, renzini17, Mowla2024}). Indeed, massive O-type stars (the stellar ionizers) are mainly forged in stellar clusters \citep[e.g.,][]{O_stars_in_clusters2012,crowther2016,massive_stars_in_YMC_2017}, especially in the most massive ones, which better sample the high mass tail of the stellar initial mass function \citep[IMF, e.g.,][]{fumagalli2011},
implying a high ionizing photons production efficiency ($\xi_{ion}$\footnote{The production rate of ionizing photon per monochromatic ultraviolet luminosity around 1500\AA: $\xi_{ion}=\frac{Q_{\rm H^o}}{L_{\rm UV}}~[s^{-1}/erg~s^{-1}~Hz^{-1}]$, where $Q_{\rm H^o}[s^{-1}] = 1.36\times 10^{12}$~L$_{\ha}[erg/s] $ \citep[][]{emami2020_xi}.}).
The inclusion of very massive stars with masses grater than 100 \msun\ in stellar evolution and atmosphere models is now becoming a key ingredient when reproducing and/or predicting the ultraviolet properties and the ionizing photon production efficiency of star-forming galaxies \citep[e.g.,][]{schaerer2024_VMS}. 

Assuming a Kroupa IMF, stellar clusters less massive than a few $10^4$~\msun\ are expected to show modest $\xi_{ion}$ values, log($\xi_{ion}$[Hz~erg$^{-1}$])$~<{25}$ (\citealt{Stanway_Eldridge2023}; see also \citealt{fumagalli2011}). 
In this view, the reionization process likely starts within the natal regions of stellar clusters, where the short-lived massive stars produce the bulk of ionizing photons. Knowing the fraction of the ionizing radiation that escapes into the intergalactic medium ($f_{\rm esc}$) and $\xi_{ion}$, one can compute the total photon rate at which a given galaxy population ionizes the IGM \citep[][]{schaerer16, robertson15}. 
The higher the $\xi_{ion}$, the lower the $f_{\rm esc}$ value that can be accommodated to sustain reionization, modulated by the integrated contribution from an arbitrarily faint population of sources \citep[e.g.,][]{atek2024Nature, simmonds2024, Harshan2024MNRAS_CANUCS}. Though still not conclusive, the identification of faint sources at $z \gtrsim 6$ showing high log($\xi_{ion})=25.85 \pm 0.05$ suggests that a modest value of $f_{\rm esc}$=5\% would be enough to reionize the Universe if such $\xi_{ion}$ is assumed valid for the whole faint population, \citet{atek2024Nature} (see also \citealt{munoz_too_many_photons2024}).

$\xi_{ion}$ positively correlates with the specific star formation rate (sSFR, e.g., \citealt{castellano2023}), which in turn correlates 
with the equivalent widths of optical lines \oiiidoublam\ and \ha\ \citep[e.g.,][]{Tang2023, caputi2024}. Recent \JWST-based findings confirm there is a high frequency of such strong optical emitters in galaxies in the reionization era at $z>5-6$ compared to cosmic noon and the local Universe \citep[][]{endsley21, endsley2023, matthee22_nircam_slitless, boyett2024}.
With a large scatter, \citet{endsley2023} also observed a slightly decreasing strength of the optical emission lines on average towards low UV luminosity galaxies.

Overall, there are (indirect) empirical pieces of evidence suggesting that star cluster formation was more vigorous in the early Universe when galaxies were denser than today 
(e.g., high star-formation per unit area, $\Sigma_{\rm SFR}$, \citealt{morishita2024_sizes, ormerod2024_sizes,matharu24, Reddy2023_keck, Reddy2023_JWST}), possibly indicating a higher gas pressure in the interstellar medium, which eventually favors the formation of star clusters (\citealt{elmegreen2018};
see also \citealt{adamo15, adamo20extreme, kruijssen12}).
The brightest and most massive star clusters are now promptly identified in high redshift lensed surveys whenever the angular resolution (both instrumental and aided by lensing) is sufficiently high. Such high spatial resolution was granted  in the pre-\JWST\ era only by Hubble \citep[][]{vanz_paving, vanz_id14, vanz19, vanz22_CFE, calura2021}. More recently, \JWST\ dramatically improved our ability to find and resolve individual star clusters \citep{adamo2024, vanzella_glass_2022, claeyssens22, adamo_sparkelr_2023, Mowla2024, vanzella2_sunrise2023}.
The behavior of $\xi_{ion}$ in the low luminosity/metallicity domain at high redshift is still under investigation, and it remains unclear below which luminosity level it (or if it) begins to dim.
Identifying ultra-faint sources is challenging at $z\gtrsim6$ also in the lensed fields, requiring relatively high magnification values ($\mu>10-20$). Once achieved, the gain at low luminosity and the enhanced spatial contrast naturally lead to approach stellar cluster luminosity-size regimes and eventually significantly low metallicity conditions \citep[e.g.,][]{vanz2023_LAP1, venditti2023}. 
In this work, we report on the serendipitous discovery of a remarkably strong yet extremely faint ionizing source lying in a poorly explored low-luminosity domain in the reionization epoch. 

Throughout this paper, we assume a flat cosmology with $\Omega_{M}$= 0.3,
$\Omega_{\Lambda}$= 0.7 and $H_{0} = 70\,{\rm km}\,{\rm s}^{-1}\,{\rm Mpc}^{-1}$. 
All magnitudes are given in the AB system \citep{Oke_1983}:
$m_{\rm AB} = 23.9 - 2.5 \log(f_\nu / \mu{\rm Jy})$.

\begin{figure}
\center
 \includegraphics[width=\columnwidth]{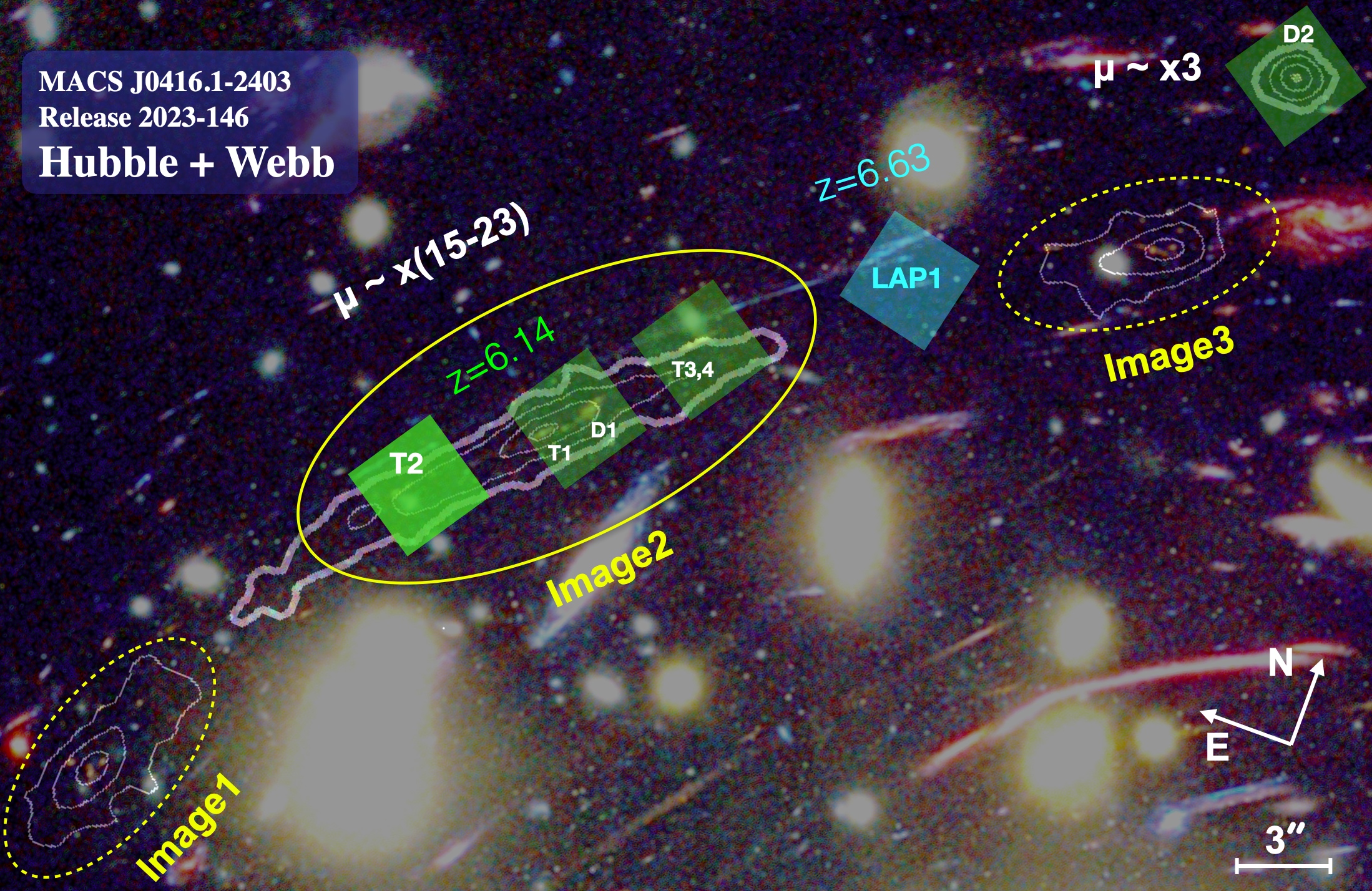}
 \caption{Schematic view of the NIRSpec IFU pointings targeting lensed dwarfs and proto-globulars at z=6.14 (transparent green boxes) and a candidate population III star complex at z=6.63, dubbed LAP1 \citep[][]{vanz2023_LAP1} (transparent cyan box). 
 Each square box resembles the IFU field of view of $3'' \times 3''$, with the white contours outlining the VLT/MUSE \lya\ emission at z=6.14 at 2-4-6-8$\sigma$ \citep[][]{vanz_mdlf}. The thick green box marks the NIRSpec pointing on source T2 discussed in this work. The background color image is the release 2023-146, which combines Hubble and \JWST\ imaging (from PEARLS team). The dotted-line ellipses ({\it Image1} and {\it Image3}) mark the two multiple images of the most magnified region marked with a larger ellipse (solid line, {\it Image2}).} 
 \label{fig:main}
\end{figure}
\begin{figure*}
\center
 \includegraphics[width=\textwidth]{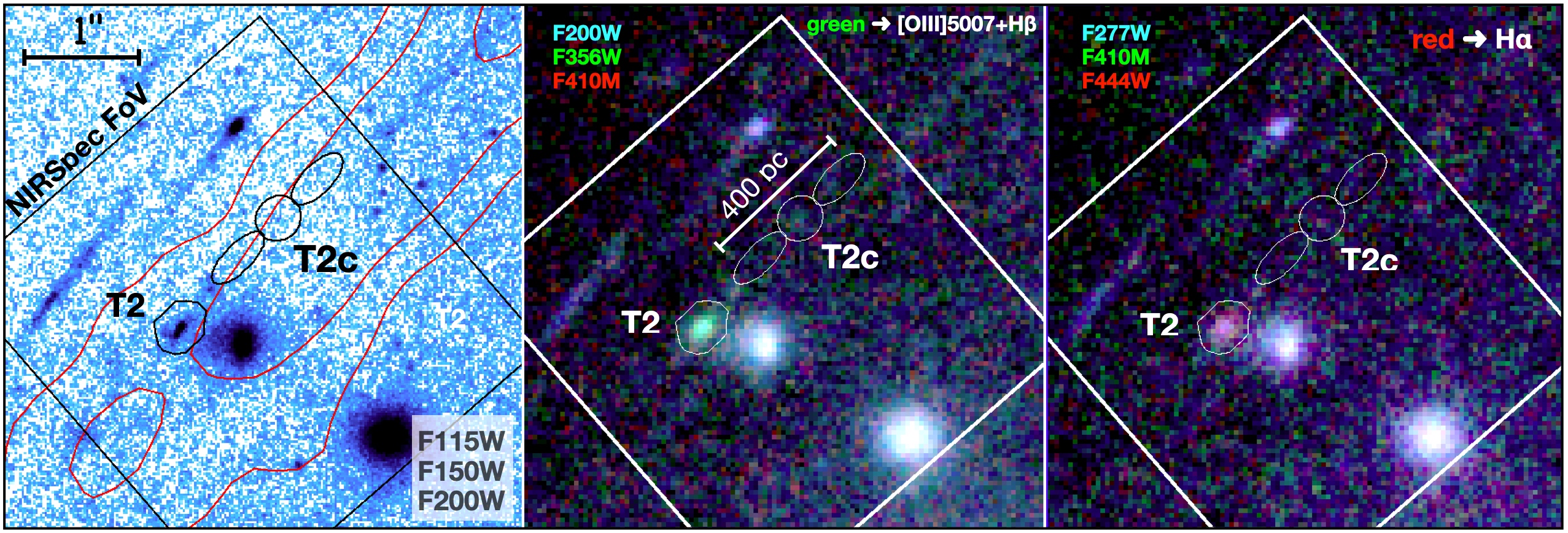}
 \caption{Overview of the \JWST/NIRSpec IFU pointing on T2. From left to right: the stacked \JWST/NIRCam F115W + F150W + F200W image with overlaid the NIRSpec IFU footprint, the location of T2 and nearby JWST-UV-dark T2c emitter, along with the VLT/MUSE \lya\ contours at 2-4-6 sigma as part of the giant \lya\ arc (red lines); the color composite image highlighting in the green channel the boost produced by \hb\ + \oiiidoublam; the color composite highlighting the presence of \ha\ in the red channel.}
 \label{fig:color}
\end{figure*}
\section{\JWST\ Oservations} \label{sec:imaging}

\subsection{\JWST\ NIRSpec/IFU }
\JWST/NIRSpec integral field unit (IFU) observations (PI Vanzella, cycle 1, prog. id 1908) were performed on October $16-17$, 2022 and August $27-31$ 2023, consisting of five pointings targeting strongly lensed dwarfs and candidate globular cluster precursors at z=6.14 (Messa et al. 2024, submitted, M24 hereafter) and a candidate population III stellar complex at redshift $z=6.63$ \citep[][]{vanz_popiii, vanz2023_LAP1}. In particular, four out of five pointings covered a lensed structure of tiny star-forming regions embedded in \lya\ nebulae at $z=6.14$ \citep[][]{vanz19,vanz_mdlf,calura2021} for a total integration time of $\simeq 18$h on target. Here, we present initial results from one of these four pointings focusing on the target T2 (see Figure~\ref{fig:main}).  

Data were reduced following the same procedures described in M24.
Briefly, we used the STScI pipeline 
\citep[v1.14.0 and 1230.pmap, ][]{bushouse23} 
and elaborated the intermediate products of stage 2 by performing customized cleaning on the eight partial cubes and combining them into the final cleaned cube. This post-processing includes background subtraction, the removal of outliers and detector defects, and the computation of the error spectrum. In addition, cross-checks on the flux calibration were performed using \JWST\ NIRCam photometry on sources lying in the same field of view (see next section). The detected sources (through their emission lines) in the reduced, post-processed, and collapsed data cube have been aligned with the \JWST/NIRCam counterparts by applying a rigid shift on RA and DEC.

\subsection{\JWST\ NIRCam}
\JWST/NIRCam observations were acquired during 2022 and 2023 as part of the two GTOs programs: the CAnadian NIRISS Unbiased Cluster Survey, CANUCS \citep[PID 1208,][]{willott2022} and the Prime Extragalactic Areas for Reionization and Lensing Science, PEARLS \citep[PID 1176,][]{windhorst23_pearls}. 
The galaxy cluster MACS~J0416 was observed in eight NIRCam filters on both programs, covering the spectral range from $0.8\mu m$ to $5\mu m$ (F090W, F115W, F150W, F200W, F277W, F356W, F410M, F444W) and combined as described in M24. Specifically, the data reduction was performed by following the prescriptions given in \citet{Yan2023_m0416}, which start from the default JWST pipeline \citep[][]{bushouse23}.
The final integration time of the stacked images is $\simeq 17,000$ seconds per filter, corresponding to 5-$\sigma$ magnitude limits for point sources of 30.1, 30.0, 30.1, 30.3 (adopting $0.1''$ diameter aperture) in the F090W, F115W, F150W, F200W short wavelength (SW) bands and 30.7, 30.8, 30.1, 30.4 ($0.2''$ diameter aperture) in the F277W, F356W, F410M, F444W long wavelength (LW) bands. The short(long) wavelength images were produced on a grid of 20(40) milliarcsec per pixel.

\begin{table*}
\caption{Observed and derived properties of T2 family.} 
\label{tab:spec}      
\centering          
\begin{tabular}{l | c c c c}     
\hline\hline  
Quantity    & T2   & T2c & T2c-tails & full-T2c \\ 
\hline
m$_{\rm UV}$, m$_{\rm opt}$ [2000, 5000\AA] & $27.60(0.10)$, $28.31(0.07)$ &$> 31.13$,  $> 30.50$ & $> 30.80$,  $> 30.15$ & $> 30.52$,  $> 29.90$\\
M$_{\rm UV}$ [2000\AA] & $-15.97\pm0.15$ &$> -12.22$ & $> -13.08$ & $> -12.80$\\
\ha\ ~~[$10^{-19} cgs$ ]      &  $15.51 \pm   0.68$ & $2.94 \pm 0.46 $ & $3.19 \pm  0.65 $& $6.03 \pm  0.86 $\\
\ha\ / \hb                     &  $2.73 \pm 0.28$   & $2.53\pm0.90$  &  $2.40\pm1.05$ &  $2.60\pm0.90$\\
\oiiiv\ ~~[$10^{-19} cgs$ ]   &  $21.40 \pm 0.77$   & $2.30 \pm 0.33$  & $2.31\pm0.58$ & $4.48 \pm 0.67$\\
EW(\ha)~~[\AA ]               &  $907 \pm 70$      & $>1300$$^{\star}$          & $>1030$          & $\gtrsim 1540$\\
EW(5007)~~[\AA ]              &  $736\pm56$         & $>600$           & $\gtrsim 430$               & $\gtrsim 670$\\
log($\xi_{ion}$~[Hz~erg$^{-1}$]) &  $25.39\pm 0.05$    & $\gtrsim26.08^\clubsuit$   & $\gtrsim25.98^\clubsuit$    & $\gtrsim26.15^\clubsuit$ \\
R3~~[$\frac{\oiiiv}{\ha/2.86}$] & $4.0\pm0.2$ & $2.5\pm0.6$ &  $2.0\pm0.6$ & $2.1\pm0.5$ \\
Z(\%), 12+log(O/H)$^{\dag}$    & $5.0^{+1.8}_{-1.3}$,~~7.4 & $2.8^{+1.7}_{-1.1}$,~~7.1   & $2.3^{+1.4}_{-1.0}$,~~$ 7.1$ & $2.4^{+1.3}_{-0.9}$,~~$7.1$ \\
$\mu$~~[tot, tang] $\ddagger$               &  18.4, 14.6   & 23.1, 17.8             & 23.1, 17.8               & 23.1, 17.8\\
\hline \hline
\end{tabular}
\tablefoot{Measured magnitudes and fluxes. De-lensed magnitudes can be derived by adding 2.5log$_{10}$($\mu_{tot}$) to the observed ones and de-lensed fluxes by dividing the observed ones by $\mu_{tot}$. The total and tangential magnification values are reported in the last raw ({\it tot} and {\it tang}) and have 10\% relative statistical error \citep[][]{bergamini_2022_J0416}. The magnitude limits are derived from the stacked F115W+F150W+F200W image (ultraviolet rest-frame, m$_{\rm UV}$) and F410M (optical rest-frame, m$_{\rm opt}$) computed at 3$\sigma$ within the apertures as indicated in Figure~\ref{fig:spec} (bottom-right panel). The reported errors on magnitudes and fluxes measurements are at $1\sigma$ confidence level, while, if not specified in the table, lower limits are reported at 3$\sigma$. Line fluxes are reported with {\it cgs} units, corresponding to erg~s$^{-1}$~cm$^{-2}$. $(\dag)$ from \citet{asplund2009}, Z$_{\odot}$ $\Rightarrow$  8.69~=~12+log(O/H). $(\star)$ Based on F410M magnitude limit only, however, after combining the F410M and F444W (removing the \ha\ flux contribution in F444W) the 3-sigma limit of the optical continuum increases to $>31$, corresponding to EW$(\ha) > 2080$\AA. ($\ddagger$) typical $1\sigma$ error of 10\%. ($\clubsuit$) the reported $3\sigma$ limits on $\xi_{ion}$ decrease of 0.22 dex when relaxing to $5\sigma$.}
\end{table*}

\section{Results}
\label{sec:discussion}

\subsection{A serendipitous extremely faint and low metallicity source at z=6.146}




Figure~\ref{fig:main} shows the \JWST/NIRSpec IFU pointing designed to target a portion of the \lya\ arc and the source dubbed T2, as part of a larger system at z=6.14 (\citealt{vanz19} and M24). T2 is detected in all NIRCam bands
and shows clear photometric excess due to \hb\ + \oiiidoublam\ (F356W, green in the rgb rendering) and \ha\ (F444W, red in the RGB rendering), see Figure~\ref{fig:color}. The same figure also shows the location of an unexpected emitter lying in the same NIRSpec/IFU field of view, arising from an object at a physical distance of $\simeq 200$ pc from T2, dubbed T2c and not detected in deep \JWST/NIRCam short wavelengths.
The T2 complex will be discussed in more detail in a forthcoming work. Here we only state that T2 shows de-lensed ultraviolet and optical magnitudes m$_{\rm UV} = 30.76$ and m$_{\rm opt}=31.47$ (magnifications and errors are reported in Table~\ref{tab:spec}), corresponding to M$_{\rm UV} = -15.97 \pm0.15$ and M$_{\rm opt} = -15.26\pm0.13$, respectively, with \ha\ and \oiiiv\ equivalent widths (EWs) of $907(\pm70)$\AA\ and $736(\pm60)$\AA. From the R3 index (= \oiiiv\ / \hb) and the calibration curves of \citet{nakajima2023} we derive a 
metallicity of 5\% solar\footnote{Adopting the high ionization conditions, as indicated by the large EW(\hb)  $\simeq 175$\AA.}, whereas from the \ha\ luminosity and the ultraviolet flux we have 
log$_{10}(\xi_{ion})=25.39\pm 0.05$ (see Table~\ref{tab:spec} in which the uncertainties are also reported).

\begin{figure*}
\center
 \includegraphics[width=\textwidth]{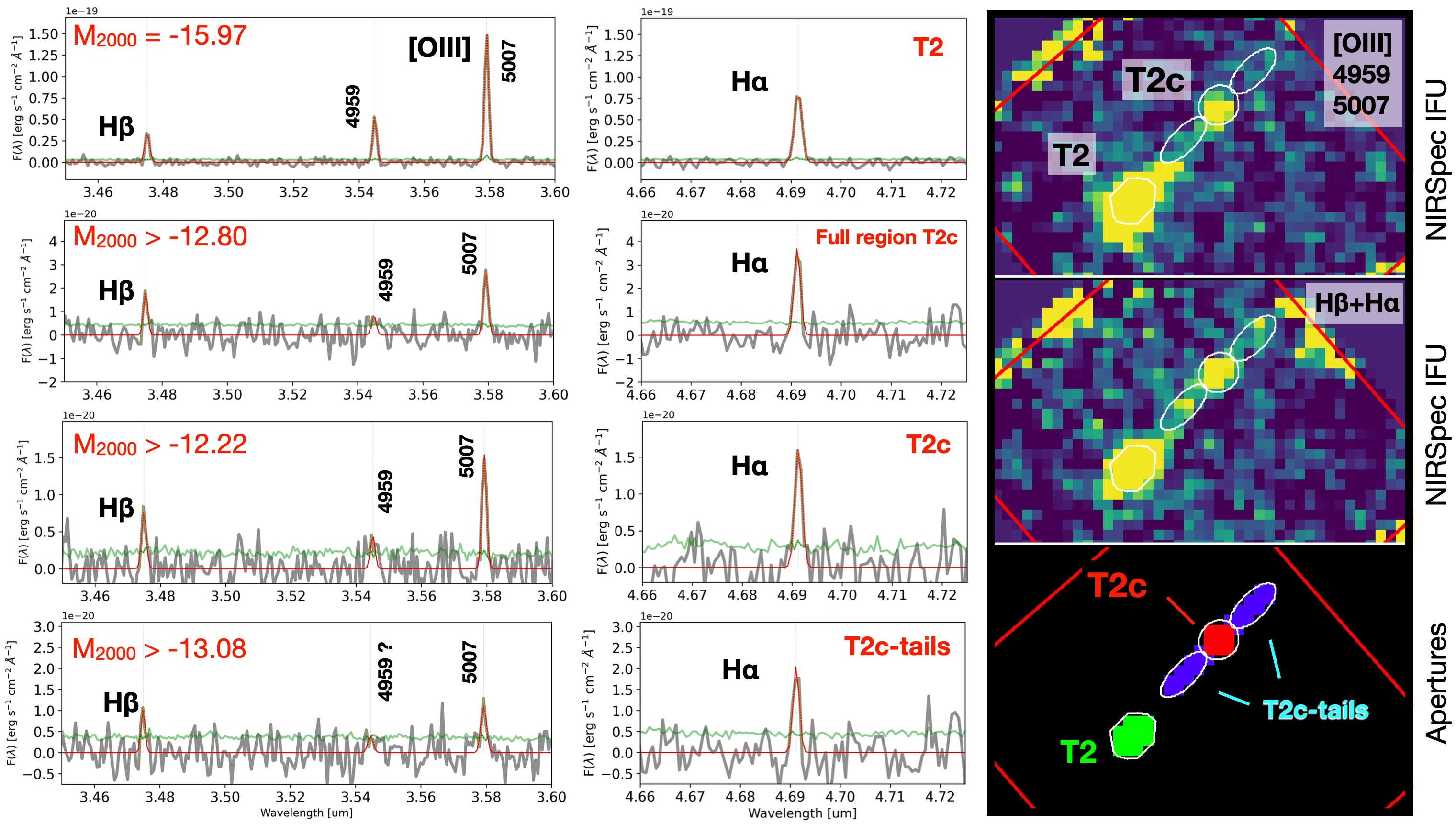}
 \caption{Left and central columns show the one-dimensional NIRSpec spectra (gray line) extracted from the masks shown in the bottom-right panel of the right-most column and labeled as T2, T2c, and T2c-tails (where T2c + T2c-tails corresponds to the ``full region T2c''). The green lines indicate the 1-sigma spectra and the red lines indicate the Gaussian fits of the emission lines. On the right, from top to bottom, are the NIRSpec \oiiidoublam\ image, the \ha\ NIRSpec image, and the masks used to extract the one-dimensional spectra.}
 \label{fig:spec}
\end{figure*}

If the source T2 belongs to a category of significantly low-luminosity objects, being among the weakest currently probed at this redshift \citep[e.g.,][]{atek2024Nature}, the nearby T2c appears to be extreme, both in terms of luminosity and the production of ionizing photons.
Two main facts emerge from T2c: 

\noindent (1) T2c is not detected in the stacked SW image (F115W+F150W+F200W, probing $\lambda \simeq 2000$\AA), down to a $3\sigma$ limit magnitude of $>31.1$, derived from the r.m.s. map on the (red) region shown in Figure~\ref{fig:spec} (see also Appendix~\ref{maglimit}). A similar lower limit is inferred from the optical continuum at $\lambda \simeq 5700$\AA, probed by the F410M medium-band filter (see also Table~\ref{tab:spec}). At the given magnification ($\times 23$), these two limits correspond to M$_{\rm UV}>-12.2$ and M$_{\rm opt}>-12.8$.
Despite relatively large uncertainties, the inferred \ha/\hb\ ratio is consistent with the value of 2.86 predicted by case B recombination theory \citep[][]{osterbrock2006}, suggesting negligible dust attenuation, as it is also further supported by the presence of the \lya-emitting region in which the sources are embedded.

\noindent (2) a nucleated \oiiidoublam\ emission is detected on T2c, surrounded by emitting \ha\ and weak oxygen (\oiiidoublam). 
We defined three regions, shown in Figure~\ref{fig:spec} (bottom-right): a peaked emission (T2c), two tails (T2c-tails), and the combination of both (full-T2c-region). The one-dimensional spectra extracted from these regions are presented in Figure~\ref{fig:spec}.
The R3 index and calibrations of \citet[][]{nakajima2023} place the metallicity of T2c at $\sim 2-3$\% solar in high ionization conditions. A similar value is inferred for T2c-tails (the spectral properties are listed in Table~\ref{tab:spec}).

%

It is worth noting that T2c is detected in the NIRCam LW bands. In particular, a green spot appears on the color image shown in Figure~\ref{fig:color}, corresponding to magnitude $30.40 \pm 0.12$ in F356W (while it is undetected in the blue, F227W, and red, F410M, channels). The NIRSpec flux of the emission lines \hb\ + \oiiidoublam\ lying in the same filter accounts for a magnitude of 30.54, implying that the detection of T2c in F356W is fully compatible with the measured line emission inferred from  NIRSpec data (the emission appears unresolved on both instruments). Similarly, the measured \ha\ flux from NIRSpec corresponds to magnitude 30.60 in F444W, which at the given depth makes T2c barely detected in the NIRCam/F444W band (SNR~$\simeq 2$).

\subsection{A very efficient ionizer}

The ultraviolet stellar continuum of T2c and the surrounding region is not detected, therefore the information on their morphology is not available. 
However, the unresolved \oiii\ emission detected on NIRCam/F356W (at S/N~$ = 9$) 
implies a size smaller than the PSF FWHM ($<0.15''$)
which along the tangential shear corresponds to a radius $< 25$ pc considering the PSF's half width at half maximum. Under the assumption that the \oiii\ emission traces the stellar component, and given its compactness, it is plausible that the dominant source of the ionizing radiation is a star cluster (or a group of clusters) confined within a few tens of parsec. Such star complexes might power an \hii\ region, shaping what we label as T2c-tails.
It is worth noting that additional undetected ionizing sources may also contribute to the ionization of the T2c tails. These are possibly indirectly traced by very weak \oiii\ emission, also detected in the T2c tails (at S/N~$\sim 3$, see Figure~\ref{fig:spec}, T2c-tails spectrum).

\begin{figure}
\center
 \includegraphics[width=\columnwidth]{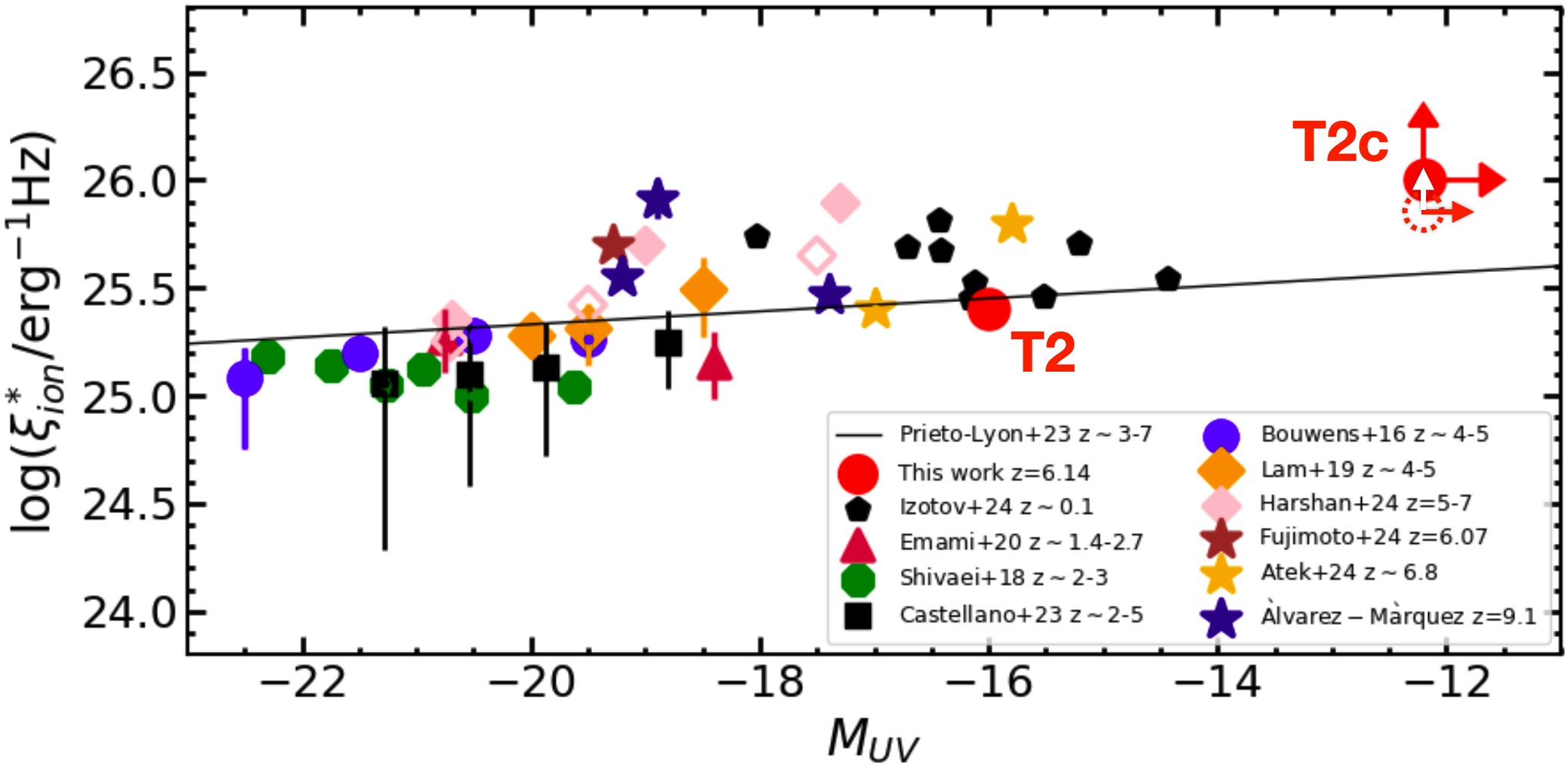}
 \caption{Collection of ionizing photon production efficiency measurements as a function of the absolute ultraviolet magnitude. The sources studied in this work extend to the lowest luminosity limits, captured in their bursty phase. The data have been retrieved from \citet{emami2020_xi, castellano2023, izotov2024_9lowZ,bouwens.xsi, Shivaei2018, Harshan2024MNRAS_CANUCS, Lam2019_xiion, atek2024Nature, Alvarez_Marquetz2024_z9p1, Fujimoto_2024_grapes} and the linear fit (solid line) from the work of \citet{Prieto-Lyon2023}. The open and solid symbols from  \citet{Harshan2024MNRAS_CANUCS} refer to NIRCam and NIRSpec-based measurements, respectively. T2c is indicated at 3 (filled red circle) and 5 (red dotted circle) sigma lower limit.}
 \label{fig:XIUV}
\end{figure}

Adopting an instantaneous burst scenario, the large EW(\ha) of T2c ($>1300$\AA\ rest-frame) implies a very young age of a few Myr. The observed absolute magnitude (M$_{\rm UV} \gtrsim -12.22$ at $3\sigma$) translates to a stellar mass lower than $2 \times 10^4$~\msun\ \citep[adopting {\tt Starburst99} models at the closest metallicity of the source,][]{leitherer14} and adopting negligible dust attenuation.  
T2c is rapidly growing caught in the very early phase of formation (less than a few Myr) with a \ha-based star formation rate of 0.032~\msun~yr$^{-1}$ \citep[][]{kenni12}, implying a sSFR~$\gtrsim 1000$~Gyr$^{-1}$.
Despite the low stellar mass, the ionizing photon production efficiency is remarkably high, log$_{10}(\xi_{ion})\gtrsim 26.08(25.86)$ at 3(5)-sigma.
Such large $\xi_{ion}$ implies massive stars are present in a still rapidly growing source, likely caught in a rare phase of evolution \citep[][]{Stanway_Eldridge2023}. The low metallicity of T2c goes into the direction of supporting the inferred large $\xi_{ion}$ values, although $\xi_{ion}$ increases slightly by $\sim 0.1$ dex when metallicity decreases from solar to 1/100 solar \citep[][]{Raiter2010}. 
It is worth noting that similar properties, though for two magnitudes brighter galaxies (and slightly more massive, $10^{5-7}$~\msun), have been reported by \citet{izotov2024_9lowZ} on a sample of nine most metal-deficient compact star-forming galaxies at $z\simeq 0.1$ (effective radii of a few tens pc in the ultraviolet). The Izotov's sample is reported in Figure~\ref{fig:XIUV} and shows relatively large $\xi_{ion}$,
log$(\xi_{ion})=25.45-25.81$.

\subsubsection{Caveats}
Noteworthy, the lower limit on log($\xi_{ion}$) close to 26 is rather challenging to reproduce for stellar evolution models unless population III stars or complexes dominated by very massive stars only are invoked (\citealt{schaerer2024_VMS}, see also \citealt{Raiter2010}). Another option that might explain such a large $\xi_{ion}$ is the pure nebular emission in the case of a spatially resolved \hii\ region, such that T2c is just fluorescing gas illuminated by ionizing radiation emitted by, e.g., the nearby source T2. This scenario would be disfavored if the stellar continuum is detected in T2c. In the present case, however, the detection of the ultraviolet continuum seems at the very limit of the current depth and non-conclusive (Appendix~\ref{maglimit}). A combination of the two, fluorescence and in-situ star formation, might also be possible. An additional caveat might be related to possible differential magnification. If the two regions emitting ultraviolet radiation and \ha\ have different sizes, then $\xi_{ion}$ would be lens-model-dependent.
However, the case of a stellar component intrinsically smaller than the \ha\ region would increase $\xi_{ion}$, being the UV part more magnified than the \ha\ region \citep[e.g.,][]{vanz_popiii}. In this work, we assume that the magnification factor does not enter in the computation of $\xi_{ion}$ (i.e., it is the same for the stellar continuum and line emission).
Last, dust attenuation might also decrease $\xi_{ion}$ (assuming $f_{\rm esc} = 0$), 
for example, an A$_{\rm UV} = 0.4$ would decrease the value of 0.16 dex; however, as discussed above, we don't have a clear indication for the presence of dust attenuation on T2c. 
In the case of ongoing in-situ star formation, it remains unclear what the real nature of T2c is in terms of underlying stellar populations. Overall, the system shows very efficient ionizing properties.

%

\section{Final Remarks}

As part of a study characterizing a wider lensed structure at z=6.14, we discovered prominent \ha\ and modest oxygen emissions in a strongly lensed source, dubbed T2c, located approximately 200 parsecs from a brighter target (T2, M$_{\rm UV}\simeq-16$). An \hii\ region of $\sim 400$ pc size is likely powered by T2c (continuum-undetected) along with possible surrounding star-forming complexes, which are also currently undetected in deep \JWST/NIRCam imaging probing the ultraviolet rest-frame (M$_{\rm UV} \gtrsim -12.22$, see Table~\ref{tab:spec}). 
In general, T2c and its local environment show extremely high ionizing photon production efficiencies, log$_{10}(\xi_{ion})\gtrsim 26$ at $3\sigma$, values which approach the maximum expected at the given metallicity (e.g., Figure~1 of \citealt{Raiter2010}).
Our results can be summarized as follows:  

\noindent (1)
\JWST/NIRSpec observations reveal the power of blind IFU spectroscopy, serendipitously confirming a super-faint (M$_{\rm UV}\gtrsim -12.22$), low stellar mass and metallicity star complex at z=6.146, dubbed T2c;

\noindent (2)
A significantly high $\xi_{ion}$ is associated with T2c, suggesting the presence of massive O-type stars in a remarkably low stellar mass object of $\lesssim 10^{4}$~\msun. In such a low-mass regime, poor sampling of the initial mass function (IMF) at high stellar masses would make such powerful ionizers statistically rare. However, numerical works \citep[e.g., review by][]{klessen2023} find that the formation of very massive stars might be favored in very low metallicity environments. Hence, stochastic IMF sampling might not be sufficient to explain the observed $\xi_{ion}$ values in such low mass regions, and changes in the physics of star formation might become important at such low regimes. When comparing T2c to the most metal-poor galaxies observed in the local Universe \citep[e.g.,][]{izotov2024_9lowZ}, it remains unclear whether T2c-like objects exist locally \citep[e.g., see also][]{Lee2009A_local_ha_uv}.

\noindent (3)
Faint ionizing and metal-poor sources (2-3\% Z$_{\odot}$) emerge around the brighter objects, suggesting that star formation can occur in very low metallicity gas conditions near already chemically evolved regions. This provides promising prospects for detecting pristine stars when gravitational lensing enhances spatial contrast. For example, the T2-T2c system is separated by approximately 35 milliarcseconds in the source plane (about the native pixel size of NIRCam, 31 milliarcsec), making studying such sources challenging without lensing.

Finally, it is worth emphasizing that a systematic search for extremely low-luminosity and metal-poor sources like T2c at these redshifts is observationally challenging. Such efforts require blind integral field spectroscopy starting from the surroundings of relatively bright and metal-poor systems \citep[e.g.,][]{venditti2023, Venditti2024}. Strong gravitational lensing significantly enhances the detection capability by increasing depth (pushing beyond magnitude 31) and spatial contrast (to a few tens of parsecs) through the amplification factor.
This approach will enable the exploration of star-forming modes in very low mass regimes, potentially reaching very low metallicity at typical star cluster scales (parsecs). To mitigate the modest field of view of NIRSpec ($3'' \times 3''$) while keeping a blind approach, deep NIRCam Wide Field Slitless Spectroscopy (WFSS) and/or deep NIRCam intermediate-band imaging will be crucial for isolating and probing rare star formation episodes under pristine conditions. 
While WFSS would require tens of hours of integration time to achieve the observed line fluxes reported here (a few $10^{-19}$ cgs, Table~\ref{tab:spec}) and has limited spectral coverage, deep intermediate-band NIRCam imaging provides a more promising and effective method for characterizing the interplay between prominent and deficient line emissions \citep[e.g.,][]{withers2023_canucs}. However, it would necessitate subsequent spectroscopic follow-up.


\begin{acknowledgements}
This work is based on observations made with the NASA/ESA/CSA 
\textit{James Webb Space Telescope} (\JWST)
and \textit{Hubble Space Telescope} (\HST). 
These observations are associated with \JWST\ GO program n.1908 (PI E. Vanzella), GTO n.1208 (CANUCS, PI C. Willot) and GTO n.1176 (PEARLS, PI R. Windhorst). We acknowledge financial support through grants PRIN-MIUR 2017WSCC32 and 2020SKSTHZ. MM acknowledges support from INAF Minigrant ``The Big-Data era of cluster lensing''. MC acknowledges support from the INAF Mini-grant ``Reionization and Fundamental Cosmology with High-Redshift Galaxies". 
EV and MM acknowledge financial support through grants PRIN-MIUR 2020SKSTHZ, the INAF GO Grant 2022 ``The revolution is around the corner: JWST will probe globular cluster precursors and Population III stellar clusters at cosmic dawn'' and by the European Union – NextGenerationEU within PRIN 2022 project n.20229YBSAN - Globular clusters in cosmological simulations and lensed fields: from their birth to the present epoch. 
RAW acknowledges support from NASA JWST Interdisciplinary Scientist grants
NAG5-12460, NNX14AN10G and 80NSSC18K0200 from GSFC. AA acknowledges support by the Swedish research council Vetenskapsr{\aa}det (2021-05559). MB and GR acknowledge support from the Slovenian national research agency ARRS through grant N1-0238. KIC acknowledges funding from the Netherlands Research School for Astronomy (NOVA) and the Dutch Research Council (NWO) through the award of the Vici Grant VI.C.212.036. FL acknowledges support from the INAF 2023 mini-grant "Exploiting the powerful capabilities of JWST/NIRSpec to unveil the distant Universe".
MG thanks the Max Planck Society for support through the Max Planck Research Group. This research has used NASA’s Astrophysics Data System, QFitsView, and SAOImageDS9, developed by Smithsonian Astrophysical Observatory.
Additionally, this work made use of the following open-source packages for Python, and we are thankful to the developers of these: Matplotlib \citep{matplotlib2007}, MPDAF \citep{MPDAF2019}, Numpy \citep[][]{NUMPY2011}. 
\end{acknowledgements}

%
%

\bibliographystyle{aa}
\bibliography{bib}

\begin{appendix} 

\section{Barely detectable of T2c}
\label{maglimit}

\begin{figure*}
\center
 \includegraphics[width=\textwidth]{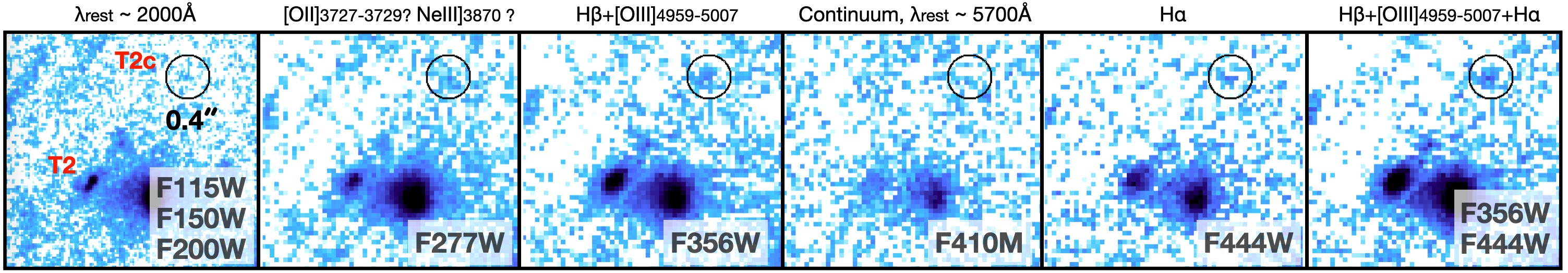}
 \caption{\JWST/NIRCam thumbnails of the region covering T2 and T2c. The black circle shows the $0.4''$ diameter aperture compatible with the aperture used on NIRSpec (red square in Figure~\ref{fig:spec}). T2c is detected in the bands enclosing rest-frame optical emission lines (confirmed with NIRSpec) and appears extremely faint in the other bands (see Table~\ref{tab:spec}). The sum of F356W and F444W provides the detection of T2c with the higher significance ($\simeq 10 \sigma$) as the result of line boosting by \hb, \oiiidoublam\ and \ha. On top are reported the most prominent lines and/or rest-frame continuum expected in the corresponding bands. The leftmost image has been slightly smoothed, adopting Gaussian with $\sigma = 0.5$ pixel.}
 \label{fig:cutouts}
\end{figure*}

Figure~\ref{fig:cutouts} shows the \JWST/NIRCam imaging centered on T2 and T2c. As already discussed in the main text T2c is detected in the bands enclosing emission lines, while it appears extremely faint or undetected in the other bands. In particular, after subtracting the local background, there is a formal 3.4 sigma detection at the position of T2c in the stacked F115W+F150W+F200W image, corresponding to magnitude $\simeq 32$ in a circular aperture of $0.1''$ diameter. It suggests a possible ultraviolet continuum detection, which is affected by severe uncertainties. For this reason, we consider a more conservative limit derived within an aperture of $0.4''$ diameter, which mimics the NIRSpec aperture used ($0.3'' \times 0.3''$) to extract the line fluxes from T2c. Such an aperture provides a lower limit of ultraviolet magnitude ($\lambda \simeq 2000$\AA\ rest-frame) $\gtrsim 31$ at 3-sigma and is the value adopted in the main text.
The F356W and F444W LW bands include the optical lines \hb+\oiiidoublam\  and \ha, respectively, which are also observed with NIRSpec and account for the detections in the same bands (as reported in Figure~\ref{maglimit}).  A possible signal is also present in F277W which might include \oiidoublam\ or \neiii\ lines, however such lines are not covered by NIRSpec and it remains unclear what is their contribution.

\end{appendix}

\end{document}